\documentclass[9pt,twocolumn,twoside]{opticajnl}
\usepackage{lineno}
\usepackage{lipsum}
\usepackage{xcolor} % in the preamble
\journal{opticajournal} % use for journal or Optica Open submissions

\setboolean{shortarticle}{true}
\title{Arbitrary Polarization Generation in Magneto-optical Metasurfaces Enabled by Bound States in the Continuum}

\author[1,*]{Siyuan Gao}
\author[2,3]{Guangtai Lu}
\author[2,3]{Satoshi iwamoto}
\author[1,*]{Yasutomo ota}

\affil[1]{Department of Applied Physics and Physico-Informatics, Keio University, Japan}
\affil[2]{Research Center for Advanced Science and Technology, The University of Tokyo, Japan}
\affil[3]{Institute of Industrial Science, The University of Tokyo, Japan}

\affil[*]{gaosy@keio.jp, ota@appi.keio.ac.jp}

\begin{abstract}

The generation of arbitrary polarization states of light is essential for optical communication and photonic information processing. Photonic crystal and metasurface platforms supporting bound states in the continuum (BICs) provide a powerful route for polarization engineering through tailoring the radiation from the resonant modes. However, existing approaches typically rely on static structural symmetry breaking or off-normal radiation, which limits continuous polarization tuning of vertical radiation.
Here, we demonstrate a magneto-optical metasurface that generates arbitrary polarization states of light at normal radiation. By applying an external magnetic field with variable orientation, a symmetry-protected BIC is transformed into a quasi-BIC whose radiation polarization can be continuously tuned. The magneto-optical perturbation drives the controlled migration of polarization singularities in momentum space, allowing the emitted states to continuously span the entire Poincar\'e sphere without structural modification. This approach establishes a compact platform for actively tunable polarization sources and polarization-encoded photonic devices.
\end{abstract}

\setboolean{displaycopyright}{false} % Do not include copyright or licensing information in submission.

\begin{document}
    %\linenumbers
    % \lipsum

\maketitle

The generation of arbitrary states of polarization (SoP) for light offers new opportunities for optical communication networks \cite{MOBICBack-LPR-communication,MOBICBack-IEEE-review}, photonic signal processing\cite{MOBICBack-PRL-signal,MOBICBack-LSc-review} and polarization-encoded information technologies\cite{MOBICBack-NC-pol-encode,MOBICBack-PRL-ecoding,MOBICBack-PRL-quantum-key,MOBICBack-Science-sensing}. Conventionally, polarization control has been achieved using birefringent materials and anisotropic optical elements\cite{MOBICBack-LSc-MS-Pol,MOBICBack-LPR-MS-Pol,MOBICBack-photoniX-anisotropic-review}. More recently, high-quality factor ($Q$) photonic platforms provide an effective route to polarization control via the manipulation of optical modes\cite{PRL-MSPolarization}. In particular, bound states in the continuum (BICs) found in periodic photonic crystals and metasurface structures have attracted considerable attention for polarization engineering\cite{BICreview,BICopticalObserv,BICinPhotoreview,BICinPhotReview2,zengManipulatingLightBound2024}. BICs often emerge at polarization singularities in momentum space, which carry integer topological charges associated with vortex-like polarization textures \cite{TopologicalNatureOptical,BICvortex,TwistedMoirePhotonic,BICvortex2,huoObservationSpatiotemporalOptical2024}. The existence of such polarization singularities enables access to a wide variety of SoP through momentum space engineering.

A widely adopted strategy to manipulate polarization singularities is geometric symmetry breaking. By reducing the symmetry of the photonic structure, symmetry-protected BICs are transformed into quasi-BICs with finite quality factors ($Q$)\cite{BIC-breakSym2024, zhangMultipleCircularPolarizations2025}, accompanied by the splitting of vortex-type polarization singularities (V points) into circular polarization singularities (C points) \cite{gorkunovMetasurfacesMaximumChirality2020,chenObservationIntrinsicChiral2023,liuCircularlyPolarizedStates2019,zengManipulatingLightBound2024}. Such symmetry-controlled approaches have enabled access to various polarization states \cite{kangCoherentFullPolarization2022,qinArbitrarilyPolarizedBound2023,huangMoireQuasiboundStates2022,Bilayer-UGR}. However, because the polarization response is directly tied to structural geometry, the achievable polarization states are fixed after fabrication, severely limiting its flexibility and reconfigurability.

To overcome the limits of fixed structure, external perturbations have been explored as a means of tuning BICs without modifying the structure\cite{tunable-graphene,yaoTunableQuasiboundStates2024,tunableBICactive}. Among them, the magneto-optical (MO) effect provides a particularly powerful approach. Previous studies have shown that MO effect can convert symmetry-protected BICs into radiative quasi-BICs \cite{yaoTunableQuasiboundStates2024,maObservationTunablePolarization2025} and induce splitting and reconfiguration of polarization singularities in momentum space\cite{zhaoMagneticallyInducedTopological2025,MO-BIC-Singularit_spliting,MO-UGR-couple,GaoACS,lvRobustGenerationIntrinsic2024a,tunableMO2025,zhaoSpinOrbitLockingChiralBound2024,tuMagneticallyTunableBound2024,tunableTHz}. However, existing approaches primarily rely on off-normal radiation or provide access only to a limited subset of polarizations, preventing continuous and deterministic control of the SoP in vertical radiation.

In this work, we demonstrate an all-dielectric MO metasurface that enables continuous and complete control of the SoP radiated from the $\Gamma$ point by manipulating the polarization singularity of the BIC mode without any structural modification. We manipulate the SoP through the azimuthal and elevation angles of the applied magnetic field, which independently control the orientation and ellipticity of the radiation. This approach establishes a high-$Q$ platform for full-Stokes polarization control, overcoming the fundamental limitations of static symmetry breaking and off-axis radiation in previous BIC-based schemes. 

\begin{figure}[ht]
\centering
\includegraphics[width=1\linewidth]{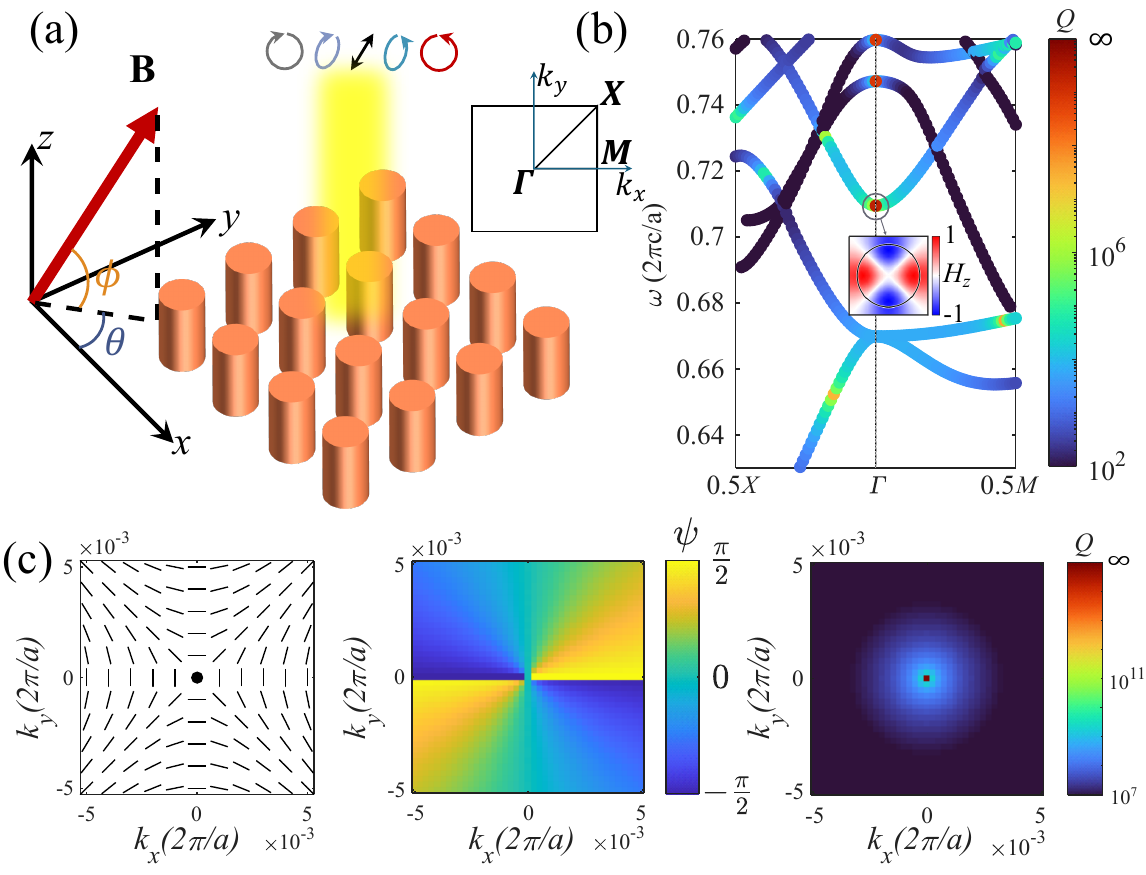}
\caption{(a) Schematic illustration of the MO metasurface and direction of external magnetic field. (b) Band diagram of the investigated structure, inset: Mode profile of the investigated BIC mode. (c) The map of polarization state, orientation angle $\psi$ and $Q$ factor of this TE band in momentum space without external magnetic field. }
\label{fig:1}
\end{figure}

%\section{Methods}
Our design is based on the all-dielectric MO metasurface schematically shown in Fig.\ref{fig:1}(a), consisting of a square array of MO nanorods suspended in air. The lattice constant is $a = 1000$ nm, the rod height $h = 1000$ nm, and the diameter $D = 800$ nm. The relative permittivity tensor of the MO material is represented as follows\cite{gxgygz}:
\begin{equation}
\hat\varepsilon = 
\begin{pmatrix}
\varepsilon_x & -ig_z & ig_y \\
ig_z & \varepsilon_y & -ig_x \\
-ig_y & ig_x & \varepsilon_z
\end{pmatrix}
\tag{1}
\label{equation1}
\end{equation}
 where the diagonal components of the permittivity tensor are taken as $\varepsilon_x = \varepsilon_y = \varepsilon_z = 2.3^2$ without loss of generality. The MO response of the material is described by the off-diagonal part of the permittivity tensor denoted as $g_i$.
% mention telecommunication band somewhere else
Assuming a magnetic field with variable orientation, the MO constants are parameterized as
\begin{equation}
\begin{aligned}
g_x = g_0\cos{\theta}\cos{\phi} \\
g_y = g_0\sin{\theta}\cos{\phi} \\
g_z = g_0\sin{\phi} \\
\end{aligned}
\tag{2}
\label{equation2}
\end{equation}
where $g_0$ denotes the overall magnitude of the MO constant. $\theta$ and $\phi$ are the azimuthal angle and the elevation angle of the applied magnetic field. Numerical simulations of the eigenmodes and radiation properties are performed using the finite-element method implemented in COMSOL multiphysics. The momentum-resolved far-field radiation is obtained by projecting the Bloch eigenmode fields onto outgoing plane-wave components\cite{TopologicalNatureOptical}.The state of polarization (SoP) is characterized using the orientation angle $\psi$ and ellipticity angle $\chi$, which are extracted from the Stokes parameters of the far-field electric field.

Firstly, we investigate the photonic band diagram in the absence of an external magnetic field ($g_0 = 0$) of the MO metasurface as illustrated in Fig. \ref{fig:1}(b). Several symmetry protected BIC modes with infinitely high $Q$ are observed at the $\Gamma$ point. The mode of interest is the polarization-degenerate TE-like quadrupole resonance mode, whose magnetic field profile is illustrated in the inset. Next, we investigate the SoP and $Q$ around the BIC point in momentum space as shown in Fig. \ref{fig:1}(c).  
A polarization vortex composed of a linearly-polarized state is observed around the $\Gamma$ point, where the orientation angle ($\psi$) undergoes a change of $-2\pi$ along a closed loop encycling the $\Gamma$ point, indicating a topological charge of $-1$ \cite{TopologicalNatureOptical}.
%at $\Gamma$ point was protected by the $C_{4V}$ symmetry of the structure. The polarization went through the change of $-2\pi$ around $\Gamma$ point, indicating a topological charge of $-1$. 
$Q$ factor diverges as the wave vector approaches the $\Gamma$ point and decreases rapidly away from this highly symmetric point in the momentum space.

%%%%%%%%%%%%%%%%%%%%%%%%%%%%%%%%%%%%%%%%%%%%%%%%%%%%%%%%%%%%%%%%%%%%%%%%%%%%%%%%%%%%%%%%%%%%%%%%%%%%%%%%%%%%%%%%%%%%%%%%%%%%%%%%%%%%%%%%%%%%%%%%%%%%%%%%%%%%%%%%%%%%%%%%%%%%%%%%%%%%%%%%%%%%%%%%%%%%%%%%%%%%%%%%%%%%%%%%%%%%%%%%%%%%%%%%%%%%%%%%%%%%%%%%%%%%%%%%%%%%%%%%%%%%%%%%%%%%%%%%%%%%%%%%%%%%%%%%%%%%%%%%%%%%%%%%%%%%%%%%%%%%%%%%%%%%%%%%%%%%%%%%%%%%%%%%%%%%%%%%
% In-plane MO effect
\begin{figure}[ht]
\centering
\includegraphics[width=\linewidth]{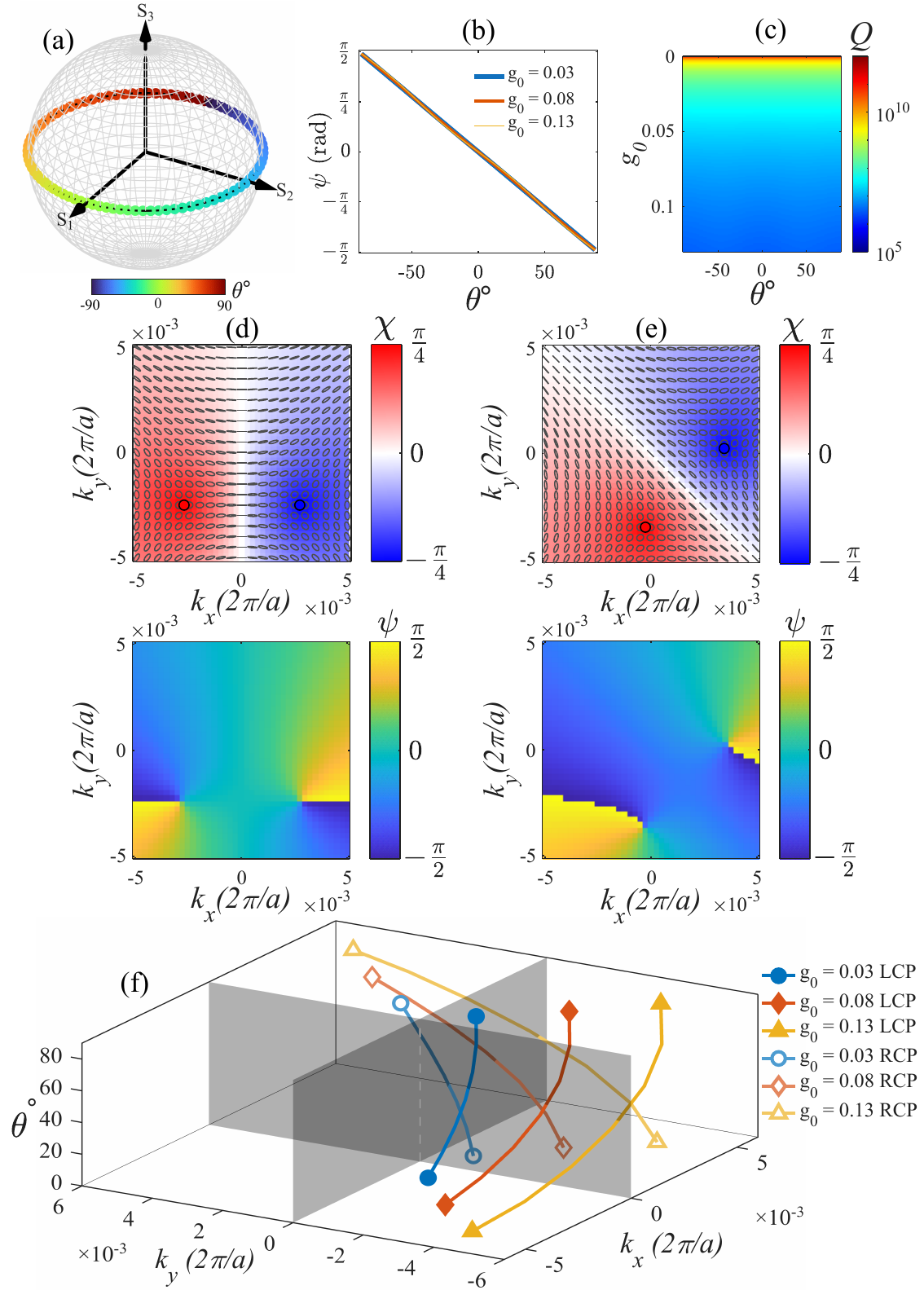}
\caption{In-plane MO effect and manipulation of SoP and $Q$ of TE-mode at $\Gamma$ point. (a) SoP at the $\Gamma$ point plotted on the Poincar\'e sphere for $g_0 = 0.03$ and $\phi = 0^\circ$ as $\theta$ varying from $-90^\circ$ to $90^\circ$, showing a continuous sweep along the equator that covers all linear-polarization orientations; (b) Orientation angle $\psi$ as a function of $\theta$ with $g_0 = 0.03, 0.08$ and $0.13$. (c) $Q$ factor distribution with varying $\theta$ and $g_0$. Polarization plot, $\chi$ and $\psi$ distribution of TE quasi-BIC mode in momentum space with (d) $g_0 = 0.08$, $\phi = 0^\circ$ and $\theta = 0^\circ$ and (e) $g_0 = 0.08$, $\phi = 0^\circ$ and $\theta = 45^\circ$; (f) The movement trajectory of LCP and RCP point with increasing $\theta$ with $g_0 = 0.03, 0.08$ and $0.13$ and $\phi = 0^\circ$.}
\label{fig:2}
\end{figure}

We next examine the impact of the in-plane ($\phi = 0^\circ$) MO effect on the far field polarization at $\Gamma$ point. The nonzero in-plane MO effect converts BIC modes to radiative modes with finite $Q$ factors \cite{yaoTunableQuasiboundStates2024}. Fig. \ref{fig:2}(a) plots the polarization states of the quasi-BIC mode at the $\Gamma$ point on the Poincar\'e sphere for $g_0 = 0.03$ as the magnetization angle $\theta$ varies from $-90^\circ$ to $90^\circ$. The resulting states are continuously distributed along the entire equatorial line of the sphere, demonstrating that by altering $\theta$, all possible orientation angle $\psi$ from $ -\frac{\pi}{2}$ to $\frac{\pi}{2}$ are generated. In addition, this indicates that the in-plane MO effect enables complete coverage of the linear polarization manifold without introducing any circular component.
The computed $\psi$ varies linearly with $\theta$ and its slope is independent of $g_0$, as shown in Fig. \ref{fig:2}(b). In contrast, the corresponding $Q$, as shown in Fig. \ref{fig:2}(c), decreases exponentially with increasing $g_0$, but remains insensitive to $\theta$. This shows that $\theta$ governs the polarization orientation $\psi$ while $g_0$ controls the radiative leakage of the quasi-BIC modes at the $\Gamma$ point. 
%The independent tunability of $\theta$ and $g_0$ enables simultaneous and continuous control over the linear polarization and $Q$, which will enable essential functionality for reconfigurable polarization devices.

To explore polarization behavior in momentum space, we map the far-field polarization distribution of the investigated TE-like eigenmodes with $g_0 = 0.08$. For $(\theta,\phi) = (0^\circ,0^\circ)$, Fig. \ref{fig:2}(d) reveals a pair of circular polarization singularities (C points) with opposite handedness located symmetrically about $k_x = 0$. The red and blue markers correspond to left (LCP) and right circular polarized (RCP) states ($\chi = \pm\frac{\pi}{4}$), respectively. 
%This C-point pair originates from the magnetically induced $C_{4V}$ symmetry breaking with preserved P symmetry in the system \cite{MO-BIC-Singularit_spliting}, which decomposes the integer charge at $\Gamma$ into two half-integer singularities. 
The colormap of the orientation angle $\psi$ exhibiting a phase winding of $-\pi$ and a topological charge of $-\frac{1}{2}$ at each C point. Along the symmetry line ($k_x=0$), the polarization remains linear with $\psi= 0$, forming an L-line that divides the two C points \cite{liuCircularlyPolarizedStates2019}.
When the magnetization is rotated to $\theta = 45^\circ$, as shown in Fig. \ref{fig:2}(e), the C points shift diagonally together with the L line, and the linear-polarization axis at $\Gamma$ rotates correspondingly to $\psi = -\frac{\pi}{4}$. The complete trajectories of the C points as a function of $\theta$ are summarized in Fig. \ref{fig:2}(f). Increasing $g_0$ enlarges the separation between the C points, while changing $\theta$ induces their azimuthal rotation around $\Gamma$ point. The in-plane magnetic field enabled the change of $\psi$ by $\theta$ and independent tuning of $Q$ by $g_0$ at the $\Gamma$ point.
%This behavior evidences a magnetically driven topological transition from a single V point (integer charge $-1$) to two C points (half charges $-\frac{1}{2}$), whose spatial trajectories can be continuously steered by the orientation and strength of the magnetic field.
%Hence, the in-plane MO effect provides a dynamic and reversible means of tailoring the topology of polarization singularities, allowing robust, real-time control over the far-field state of polarization in quasi-BIC metasurfaces.

\begin{figure}[ht]
\centering
\includegraphics[width=\linewidth]{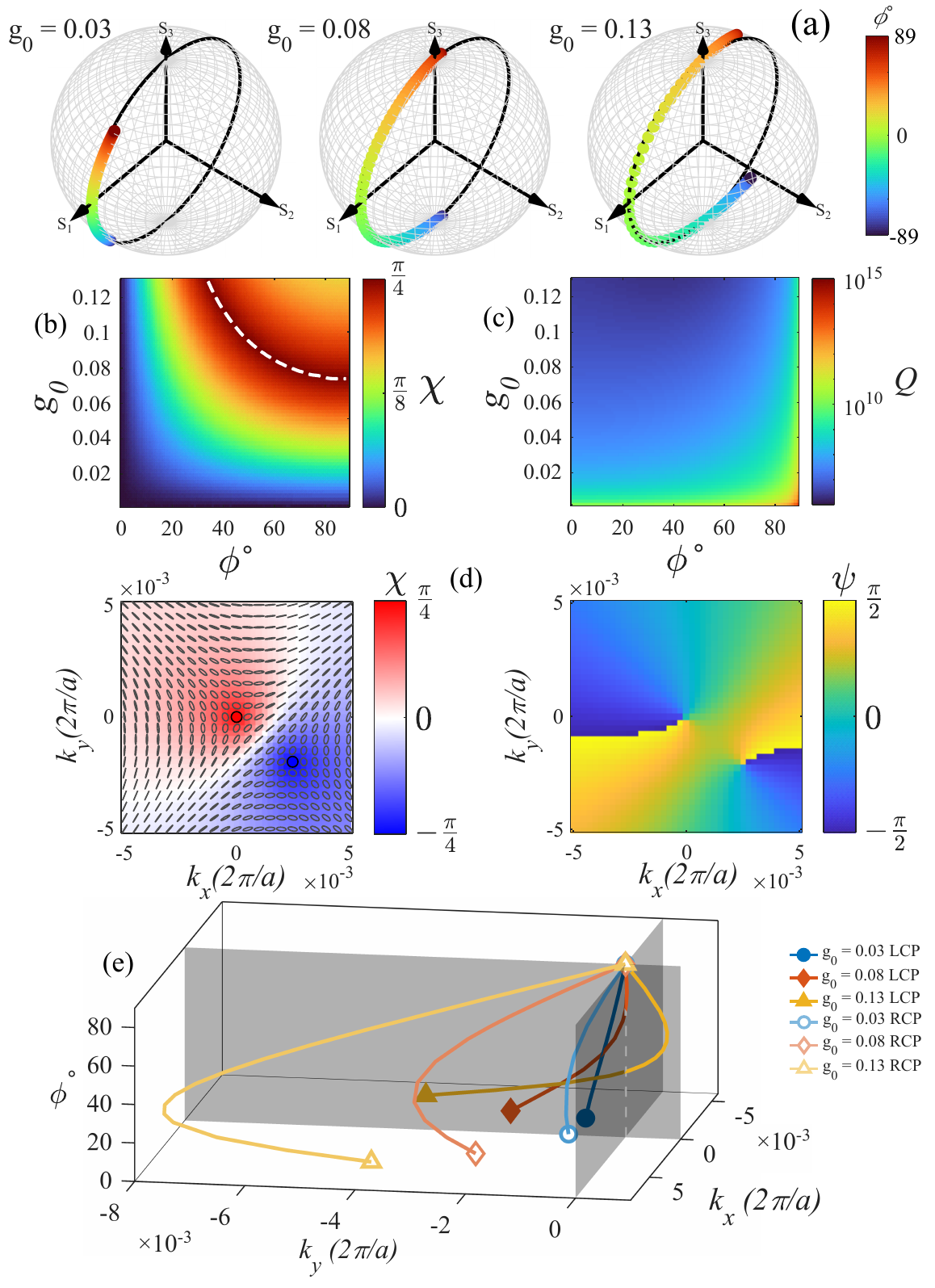}
\caption{Out-of-plane MO effect and manipulation of SoP and $Q$ of TE-mode at $\Gamma$ point. (a) SoP at the $\Gamma$ point plotted on the Poincar\'e sphere for $g_0 = 0.03, 0.08$ and $0.13$ as $\phi$ varying from $-89^\circ$ to $89^\circ$; (b) Ellipticity $\chi$ distribution with varying  $\phi$ ($0^\circ  - 89^\circ$) and $g_0$, the white dashed line is the iso-valued line presenting $\chi = \frac{\pi}{4}$; (c) $Q$ factor distribution with varying $\phi$ ($0^\circ -89^\circ$) and $g_0$; (d) Polarization plot, $\chi$ and $\psi$ distribution in momentum space with $g_0 = 0.08$, $\phi = 66.1^\circ$ and $\theta = 0^\circ$; (e) The trajectory of LCP and RCP point with increasing $\phi$ with $g_0 = 0.03, 0.08, 0.13$ under $\theta = 0^\circ$.}
\label{fig:3}
\end{figure}
%% start and end point of the C points
%It is required how this motion will result in the control over polarization at Gamma point.

%%%%%%%%%%%%%%%%%%%%%%%%%%%%%%%%%%%%%%%%%%%%%%%%%%%%%%%%%%%%%%%%%%%%%%%%%%%%%%%%%%%%%%%%%%%%%%%%%%%%%%%%%%%%%%%%%%%%%%%%%%%%%%%%%%%%%%%%%%%%%%%%
%% out-of-plane MO effect
The effect of an out-of-plane magnetic field ($\phi \neq 0^\circ$) is investigated in Fig.~\ref{fig:3}. 
Fig.~\ref{fig:3}(a) compares the evolution of the state of polarization (SoP) at the $\Gamma$ point under varying elevation angle $\phi$ for $g_0 = 0.03$, $0.08$ and $0.13$, with the azimuthal angle fixed at $\theta = 0^\circ$. The SoPs are plotted on the Poincar\'e sphere, where the color scale denotes $\phi$ ranging from $-89^\circ$ to $89^\circ$. The case $\phi = 90^\circ$ is excluded, as it corresponds to a purely out-of-plane magnetization ($g_x = g_y = 0$), under which the symmetry-protected BIC at the $\Gamma$ point is restored.
For a weak MO constant ($g_0 = 0.03$), the SoP follows a near-vertical trajectory along the $S_2 = 0$ meridian, where the ellipticity $\chi$ increases gradually with increasing $|\phi|$. 
As $g_0$ increases to $0.08$, the trajectory extends toward the north and south poles ($S_3 = \pm 1$), reaching pure LCP and RCP states at $\phi = \pm 66.1^\circ$. Further increasing $g_0$ to $0.13$ causes the trajectory to cross the poles at $\phi = \pm 34^\circ$ and continue into the opposite hemisphere. These results indicate that $\phi$ continuously controls the ellipticity through the $S_3$ component, while $g_0$ determines the accessible range of polarization states.
Fig.~\ref{fig:3}(b) presents the ellipticity $\chi$ of the quasi-BIC mode at the $\Gamma$ point as a function of $\phi$ and $g_0$. 
The ellipticity $\chi$ evolves continuously in the $(g_0,\phi)$ parameter space.
The white dashed curve denotes the iso-contour $\chi =\frac{\pi}{4}$, corresponding to a pure circular polarization state. To realize the pure circular state, the MO constant $g_0$ must be larger than $0.074$.
%indicating that pure circular polarization is accessed at specific parameter combinations rather than over a finite region
%Fig.~\ref{fig:3}(c) shows the corresponding orientation angle $\psi$ at the $\Gamma$ point. As the polarization approaches the circular limit ($\chi \rightarrow \frac{\pi}{4}$), $\psi$ varies slightly from $0$ to $0.0124\pi$. Upon reaching pure circular polarization, $\psi$ exhibits an apparent discontinuity and takes the value of $-\frac{\pi}{2}$ as the trajectory continues into the opposite hemisphere ($S_2<0$) on the Poincar\'e sphere (Fig. \ref{fig:3}(a)). 
Fig. \ref{fig:3}(c) illustrates the corresponding $Q$ of the QBIC mode at the $\Gamma$ point as a function of $g_0$ and $\phi$. The $Q$ decreases with increasing $g_0$ or decreasing $\phi$, reflecting enhanced radiative loss induced by symmetry breaking due to the MO effect. When $\phi$ reached $90^\circ$, the $Q$ factor diverges again, indicating recovery of the symmetry-protected BIC at the $\Gamma$ point.
%This discontinuity arises because the orientation angle $\psi$ becomes ill-defined at pure circular polarization, which signifies the formation of a polarization singularity, namely a C point. 

The topological interpretation of pure circular SoP is confirmed in Fig.~\ref{fig:3}(d) for $g_0 = 0.08$ and $\phi = 66.1^\circ$, where the SoP map in momentum space shows that $\Gamma$ point exhibits pure circular polarization ($\chi = \frac{\pi}{4}$) with half topological charge, while an opposite-handed C point appears at an off-$\Gamma$ wave vector $(k_x,k_y)=(0.005\frac{\pi}{a},-0.004\frac{\pi}{a})$. 
The corresponding orientation map reveals half-integer topological charges associated with these points.
%%%%%
Fig.~\ref{fig:3}(e) traces the full trajectories of the C points as $\phi$ increases from $0^\circ$ to $90^\circ$. 
For a weak MO coupling strength ($g_0 = 0.03$), the LCP and RCP points approach each other and annihilate at the $\Gamma$ point only when $\phi = 90^\circ$. 
For stronger coupling conditions ($g_0 = 0.08$ and $0.13$), one of the C points crosses the $\Gamma$ point at $\phi < 90^\circ$, giving rise to circularly polarized radiation with finite $Q$, before both C points annihilate again at $\phi = 90^\circ$. 
These results demonstrate that the emergence of circular polarization at the $\Gamma$ point is governed by the out-of-plane component of the external magnetic field.

\begin{figure}[ht]
\centering
\includegraphics[width=1\linewidth]{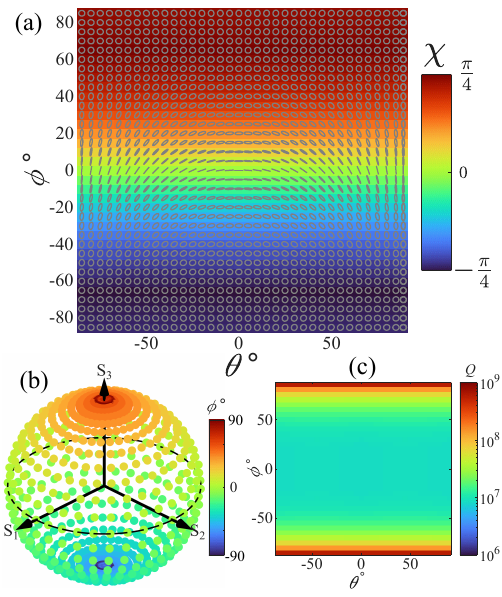}
\caption{Arbitrary polarization states control. The SoP at $\Gamma$ point under different $\theta$ and $\phi$, where polarization plot and colormap of ellipticity $\chi$ (rad) (a), the SOP plot on Poincar\'e sphere and the corresponding $Q$ factor (c).}
\label{fig:4}
\end{figure}

Finally, we demonstrate the generation of arbitrary SoP of the quasi-BIC mode at the $\Gamma$ point by jointly tuning $\theta$ and $\phi$ under a fixed $g_0=0.08$. The parameters were swept over $\theta = [-90^\circ,90^\circ]$ and $\phi = [-89^\circ,89^\circ]$
The resulting polarization states, shown in Fig. \ref{fig:4}(a), reveal that tuning $\theta$ continuously rotates the linear polarization orientation, while varying $\phi$ introduces and controls the circular component of the polarization. The $\chi = \frac{\pi}{4}$ is realized at $\phi = 66.1^\circ$. By combining both degrees of freedom, the emitted SoPs can be tuned across all possible elliptical states, enabling full coverage of the Poincar\'e sphere as confirmed in Fig. \ref{fig:4}(b).
This demonstrates that the designed MO metasurface supports complete and continuous generation of arbitrary polarization states, from linear to elliptical to circular, using only the variation of the applied magnetic field.
The corresponding map of $Q$ at $\Gamma$ point, as shown in Fig. \ref{fig:4}(c), which can be tuned by changing the value of $g_0$.
These results establish a magnetically reconfigurable all-dielectric metasurface capable of deterministic and lossless polarization control at a single-frequency BIC resonance.

%\section{Conclusions}
% Our simulations reveal that the application of the external magnetic field lifts the degeneracy of the BIC mode, converting it into a radiative quasi-BIC mode whose polarization state is governed by the field’s orientation. By systematically sweeping the azimuthal and elevation angles, we achieve continuous tuning of the far-field polarization and ellipticity across nearly the entire Poincar\'e sphere. These results establish a versatile and effective strategy for dynamic polarization control in photonic metasurfaces, providing a compact platform for advanced applications such as reconfigurable polarization sources and tunable nonreciprocal optical devices.
In summary, our simulations demonstrate that the application of an external magnetic field transforms the BIC mode into a radiative quasi-BIC mode whose polarization can be deterministically controlled solely by the magnetic field orientation.
We identified that the elevation angle ($\phi$) controls the polarization orientation ($\psi$) and the azimuthal angle ($\theta$) controls the ellipticity ($\chi$), while the strength of the MO effect ($g_0$) helps control $Q$ factor. 
This work establishes a magneto-optical route toward all-dielectric metasurfaces capable of full Stokes-parameter control without structural modification.
The demonstrated mechanism provides a compact and reconfigurable platform for next-generation photonic technologies.

\begin{backmatter}
\bmsection{Funding} This work was supported by JST FOREST (JPMJFR213F), JST CREST (JPMJCR19T1) and KAKENHI (25K01697, 24K17582), Iketani Foundation, Nippon Sheet Glass Foundation. 

% \bmsection{Acknowledgment} We thank the fruitful discussion with 

% \bmsection{Disclosures} Disclosures should be listed in a separate section at the end of the manuscript. List the Disclosures codes identified on the \href{https://opg.optica.org/submit/review/conflicts-interest-policy.cfm}{Conflict of Interest policy page}. If there are no disclosures, then list ``The authors declare no conflicts of interest.''

% \smallskip

% \noindent Here are examples of disclosures:

% \bmsection{Disclosures} ABC: 123 Corporation (I,E,P), DEF: 456 Corporation (R,S). GHI: 789 Corporation (C).

\bmsection{Disclosures} The authors declare no conflicts of interest.

\end{backmatter}

% \section{References}

% %Note that \emph{Optics Letters} and \emph{Optica} short articles use an abbreviated reference style. Citations to journal articles should omit the article title and final page number; this abbreviated reference style is produced automatically when the \emph{Optics Letters} journal option is selected in the template, if you are using a .bib file for your references.

% \bigskip
% %\noindent Add citations manually or use BibTeX. See %\cite{Zhang:14,OPTICA,FORSTER2007,testthesis,manga_rao_single_2007}. List up to three author names in references. If there are more than three authors, list the first three followed by \emph{et al.}

% Bibliography
\bibliography{sample}

@article{Bilayer-UGR,
  title = {Three-{{Dimensional Reconfigurable Optical Singularities}} in {{Bilayer Photonic Crystals}}},
  author = {Ni, Xueqi and Liu, Yuan and Lou, Beicheng and Zhang, Mingjie and Hu, Evelyn L. and Fan, Shanhui and Mazur, Eric and Tang, Haoning},
  year = {2024},
  month = feb,
  journal = {Physical Review Letters},
  volume = {132},
  number = {7},
  pages = {073804},
  issn = {0031-9007, 1079-7114},
  doi = {10.1103/PhysRevLett.132.073804},
  urldate = {2025-06-28},
  langid = {english}
}

@article{chenObservationIntrinsicChiral2023,
  title = {Observation of Intrinsic Chiral Bound States in the Continuum},
  author = {Chen, Yang and Deng, Huachun and Sha, Xinbo and Chen, Weijin and Wang, Ruize and Chen, Yu-Hang and Wu, Dong and Chu, Jiaru and Kivshar, Yuri S. and Xiao, Shumin and Qiu, Cheng-Wei},
  year = {2023},
  month = jan,
  journal = {Nature},
  volume = {613},
  number = {7944},
  pages = {474--478},
  issn = {0028-0836, 1476-4687},
  doi = {10.1038/s41586-022-05467-6},
  urldate = {2024-12-19},
  langid = {english}
}

@article{gorkunovMetasurfacesMaximumChirality2020,
  title = {Metasurfaces with {{Maximum Chirality Empowered}} by {{Bound States}} in the {{Continuum}}},
  author = {Gorkunov, Maxim V. and Antonov, Alexander A. and Kivshar, Yuri S.},
  year = {2020},
  month = aug,
  journal = {Physical Review Letters},
  volume = {125},
  number = {9},
  pages = {093903},
  issn = {0031-9007, 1079-7114},
  doi = {10.1103/PhysRevLett.125.093903},
  urldate = {2025-06-28},
  langid = {english}
}

@article{huangMoireQuasiboundStates2022,
  title = {Moir{\'e} {{Quasibound States}} in the {{Continuum}}},
  author = {Huang, Lei and Zhang, Weixuan and Zhang, Xiangdong},
  year = {2022},
  month = jun,
  journal = {Physical Review Letters},
  volume = {128},
  number = {25},
  pages = {253901},
  issn = {0031-9007, 1079-7114},
  doi = {10.1103/PhysRevLett.128.253901},
  urldate = {2024-12-19},
  langid = {english}
}

@article{huoObservationSpatiotemporalOptical2024,
  title = {Observation of Spatiotemporal Optical Vortices Enabled by Symmetry-Breaking Slanted Nanograting},
  author = {Huo, Pengcheng and Chen, Wei and Zhang, Zixuan and Zhang, Yanzeng and Liu, Mingze and Lin, Peicheng and Zhang, Hui and Chen, Zhaoxian and Lezec, Henri and Zhu, Wenqi and Agrawal, Amit and Peng, Chao and Lu, Yanqing and Xu, Ting},
  year = {2024},
  month = apr,
  journal = {Nature Communications},
  volume = {15},
  number = {1},
  pages = {3055},
  issn = {2041-1723},
  doi = {10.1038/s41467-024-47475-2},
  urldate = {2025-02-11},
  langid = {english}
}

@article{kangCoherentFullPolarization2022,
  title = {Coherent Full Polarization Control Based on Bound States in the Continuum},
  author = {Kang, Ming and Zhang, Ziying and Wu, Tong and Zhang, Xueqian and Xu, Quan and Krasnok, Alex and Han, Jiaguang and Al{\`u}, Andrea},
  year = {2022},
  month = aug,
  journal = {Nature Communications},
  volume = {13},
  number = {1},
  pages = {4536},
  issn = {2041-1723},
  doi = {10.1038/s41467-022-31726-1},
  urldate = {2025-06-28},
  langid = {english}
}

@article{liuCircularlyPolarizedStates2019,
  title = {Circularly {{Polarized States Spawning}} from {{Bound States}} in the {{Continuum}}},
  author = {Liu, Wenzhe and Wang, Bo and Zhang, Yiwen and Wang, Jiajun and Zhao, Maoxiong and Guan, Fang and Liu, Xiaohan and Shi, Lei and Zi, Jian},
  year = {2019},
  month = sep,
  journal = {Physical Review Letters},
  volume = {123},
  number = {11},
  pages = {116104},
  issn = {0031-9007, 1079-7114},
  doi = {10.1103/PhysRevLett.123.116104},
  urldate = {2024-12-19},
  langid = {english}
}

@article{lvRobustGenerationIntrinsic2024a,
  title = {Robust Generation of Intrinsic {{C}} Points with Magneto-Optical Bound States in the Continuum},
  author = {Lv, Wenjing and Qin, Haoye and Su, Zengping and Zhang, Chengzhi and Huang, Jiongpeng and Shi, Yuzhi and Li, Bo and Genevet, Patrice and Song, Qinghua},
  year = {2024},
  month = nov,
  journal = {Science Advances},
  volume = {10},
  number = {46},
  pages = {eads0157},
  issn = {2375-2548},
  doi = {10.1126/sciadv.ads0157},
  urldate = {2024-12-19},
  langid = {english}
}

@article{MO-BIC-Singularit_spliting,
  title = {Magnetic Modulation of Topological Polarization Singularities in Momentum Space},
  author = {Zhao, Chen and Dong, Shaohua and Zhang, Qing and Zeng, Yixuan and Hu, Guangwei and Zhang, Yongzhe},
  year = {2022},
  month = jun,
  journal = {Optics Letters},
  volume = {47},
  number = {11},
  pages = {2754},
  issn = {0146-9592, 1539-4794},
  doi = {10.1364/OL.458285},
  urldate = {2025-06-28},
  langid = {english}
}

@article{MO-UGR-couple,
  title = {Dynamical Control of Topological Unidirectional Guided Resonances via External Magnetic Field},
  author = {Zheng, Bao Jie and Shi, Wei Jie and Dong, Hui Yuan and Li, Yong Tao and Li, Jia Qi and Dong, Zheng-Gao and Wang, Jin},
  year = {2025},
  month = jan,
  journal = {Physical Review Research},
  volume = {7},
  number = {1},
  pages = {013091},
  issn = {2643-1564},
  doi = {10.1103/PhysRevResearch.7.013091},
  urldate = {2025-02-28},
  langid = {english}
}

@article{qinArbitrarilyPolarizedBound2023,
  title = {Arbitrarily Polarized Bound States in the Continuum with Twisted Photonic Crystal Slabs},
  author = {Qin, Haoye and Su, Zengping and Liu, Mengqi and Zeng, Yixuan and Tang, Man-Chung and Li, Mengyao and Shi, Yuzhi and Huang, Wei and Qiu, Cheng-Wei and Song, Qinghua},
  year = {2023},
  month = mar,
  journal = {Light: Science \& Applications},
  volume = {12},
  number = {1},
  pages = {66},
  issn = {2047-7538},
  doi = {10.1038/s41377-023-01090-w},
  urldate = {2024-12-19},
  langid = {english}
}

@article{BIC-breakSym2024,
  title = {Tailoring Topological Nature of Merging Bound States in the Continuum by Manipulating Structure Symmetry of the All-Dielectric Metasurface},
  author = {Sun, Guangcheng and Wang, Yue and Li, Yaohe and Cui, Zijian and Chen, Wenshuo and Zhang, Kuang},
  year = {2024},
  month = jan,
  journal = {Physical Review B},
  volume = {109},
  number = {3},
  pages = {035406},
  issn = {2469-9950, 2469-9969},
  doi = {10.1103/PhysRevB.109.035406},
  urldate = {2024-12-19},
  langid = {english}
}

@article{TopologicalNatureOptical,
  title = {Topological {{Nature}} of {{Optical Bound States}} in the {{Continuum}}},
  author = {Zhen, Bo and Hsu, Chia Wei and Lu, Ling and Stone, A. Douglas and Solja{\v c}i{\'c}, Marin},
  year = {2014},
  month = dec,
  journal = {Physical Review Letters},
  volume = {113},
  number = {25},
  pages = {257401},
  issn = {0031-9007, 1079-7114},
  doi = {10.1103/PhysRevLett.113.257401},
  urldate = {2025-01-24},
  copyright = {http://link.aps.org/licenses/aps-default-license},
  langid = {english}
}

@article{tuMagneticallyTunableBound2024,
  title = {Magnetically Tunable Bound States in the Continuum with Arbitrary Polarization and Intrinsic Chirality},
  author = {Tu, Qing-An and Zhou, Hongxin and Zhao, Dong and Meng, Yan and Gong, Maohua and Gao, Zhen},
  year = {2024},
  month = dec,
  journal = {Photonics Research},
  volume = {12},
  number = {12},
  pages = {2972},
  issn = {2327-9125},
  doi = {10.1364/PRJ.539830},
  urldate = {2025-01-07},
  langid = {english}
}

@article{yaoTunableQuasiboundStates2024,
  title = {Tunable Quasi-Bound States in the Continuum in Magneto-Optical Metasurfaces},
  author = {Yao, Enxu and Su, Zhaoxian and Bi, Yu and Wang, Yongtian and Huang, Lingling},
  year = {2024},
  month = sep,
  journal = {Journal of Physics D: Applied Physics},
  volume = {57},
  number = {37},
  pages = {375104},
  issn = {0022-3727, 1361-6463},
  doi = {10.1088/1361-6463/ad5215},
  urldate = {2024-12-19},
  langid = {english}
}

@article{zengManipulatingLightBound2024,
  title = {Manipulating {{Light}} with {{Bound States}} in the {{Continuum}}: From {{Passive}} to {{Active Systems}}},
  shorttitle = {Manipulating {{Light}} with {{Bound States}} in the {{Continuum}}},
  author = {Zeng, Yixuan and Zhang, Xudong and Ouyang, Xu and Li, Yingjie and Qiu, Cheng-Wei and Song, Qinghai and Xiao, Shumin},
  year = {2024},
  month = sep,
  journal = {Advanced Optical Materials},
  volume = {12},
  number = {25},
  pages = {2400296},
  issn = {2195-1071, 2195-1071},
  doi = {10.1002/adom.202400296},
  urldate = {2025-06-28},
  langid = {english}
}

@article{zhangMultipleCircularPolarizations2025,
  title = {Multiple {{Circular Polarizations Coexisting}} with {{Bound States}} in the {{Continuum Without Breaking Symmetry}}},
  author = {Zhang, Xiao and Xu, JiPeng and Zhu, ZhiHong},
  year = {2025},
  month = feb,
  journal = {Laser \& Photonics Reviews},
  volume = {19},
  number = {3},
  pages = {2401138},
  issn = {1863-8880, 1863-8899},
  doi = {10.1002/lpor.202401138},
  urldate = {2025-06-28},
  langid = {english}
}

@article{zhaoSpinOrbitLockingChiralBound2024,
  title = {Spin-{{Orbit-Locking Chiral Bound States}} in the {{Continuum}}},
  author = {Zhao, Xingqi and Wang, Jiajun and Liu, Wenzhe and Che, Zhiyuan and Wang, Xinhao and Chan, C. T. and Shi, Lei and Zi, Jian},
  year = {2024},
  month = jul,
  journal = {Physical Review Letters},
  volume = {133},
  number = {3},
  pages = {036201},
  issn = {0031-9007, 1079-7114},
  doi = {10.1103/PhysRevLett.133.036201},
  urldate = {2024-12-19},
  langid = {english}
}

@article{tunableTHz,
  title = {Observation of Tunable Polarization States Evolution in Bound States in the Continuum Based on Terahertz Metasurfaces},
  author = {Ma, Yumeng and Deng, Fangze and Ma, Ke and Hou, Xiang and Han, Zhihua and Li, Yuchao and Cheng, Keke and Shao, Yansheng and Wang, Chenglong and Liu, Meng and Zhang, Huiyun and Zhang, Yuping},
  year = {2025},
  month = aug,
  journal = {Journal of Applied Physics},
  volume = {138},
  number = {5},
  pages = {053101},
  issn = {0021-8979, 1089-7550},
  doi = {10.1063/5.0282994},
  urldate = {2025-08-08},
  langid = {english}
}

@article{BICinPhotoreview,
  title = {Applications of Bound States in the Continuum in Photonics},
  author = {Kang, Meng and Liu, Tao and Chan, C. T. and Xiao, Meng},
  year = {2023},
  month = nov,
  journal = {Nature Reviews Physics},
  volume = {5},
  number = {11},
  pages = {659--678},
  issn = {2522-5820},
  doi = {10.1038/s42254-023-00642-8}
}

@article{BICinPhotReview2,
  title = {Optical Bound States in the Continuum in Periodic Structures: Mechanisms, Effects, and Applications},
  shorttitle = {Optical Bound States in the Continuum in Periodic Structures},
  author = {Wang, Jiajun and Li, Peishen and Zhao, Xingqi and Qian, Zhiyuan and Wang, Xinhao and Wang, Feifan and Zhou, Xinyi and Han, Dezhuan and Peng, Chao and Shi, Lei and Zi, Jian},
  year = {2024},
  journal = {Photonics Insights},
  volume = {3},
  number = {1},
  pages = {R01},
  issn = {2791-1748},
  doi = {10.3788/PI.2024.R01},
  urldate = {2025-08-11},
  langid = {english}
}

@article{BICopticalObserv,
  title = {Observation of Trapped Light within the Radiation Continuum},
  author = {Hsu, Chia Wei and Zhen, Bo and Lee, Jeongwon and Chua, Song-Liang and Johnson, Steven G. and Joannopoulos, John D. and Solja{\v c}i{\'c}, Marin},
  year = {2013},
  month = jul,
  journal = {Nature},
  volume = {499},
  number = {7457},
  pages = {188--191},
  issn = {0028-0836, 1476-4687},
  doi = {10.1038/nature12289},
  urldate = {2025-08-08},
  copyright = {http://www.springer.com/tdm},
  langid = {english}
}

@article{BICreview,
  title = {Bound States in the Continuum},
  author = {Hsu, Chia Wei and Zhen, Bo and Stone, A. Douglas and Joannopoulos, John D. and Solja{\v c}i{\'c}, Marin},
  year = {2016},
  month = jul,
  journal = {Nature Reviews Materials},
  volume = {1},
  number = {9},
  publisher = {{Springer Science and Business Media LLC}},
  issn = {2058-8437},
  doi = {10.1038/natrevmats.2016.48},
  urldate = {2025-07-18},
  copyright = {https://www.springernature.com/gp/researchers/text-and-data-mining},
  langid = {english}
}

@article{BICvortex,
  title = {Observation of {{Polarization Vortices}} in {{Momentum Space}}},
  author = {Zhang, Yiwen and Chen, Ang and Liu, Wenzhe and Hsu, Chia Wei and Wang, Bo and Guan, Fang and Liu, Xiaohan and Shi, Lei and Lu, Ling and Zi, Jian},
  year = {2018},
  month = may,
  journal = {Physical Review Letters},
  volume = {120},
  number = {18},
  pages = {186103},
  issn = {0031-9007, 1079-7114},
  doi = {10.1103/PhysRevLett.120.186103},
  urldate = {2025-08-11},
  langid = {english}
}

@article{BICvortex2,
  title = {Generating Optical Vortex Beams by Momentum-Space Polarization Vortices Centred at Bound States in the Continuum},
  author = {Wang, Bo and Liu, Wenzhe and Zhao, Maoxiong and Wang, Jiajun and Zhang, Yiwen and Chen, Ang and Guan, Fang and Liu, Xiaohan and Shi, Lei and Zi, Jian},
  year = {2020},
  month = oct,
  journal = {Nature Photonics},
  volume = {14},
  number = {10},
  pages = {623--628},
  issn = {1749-4885, 1749-4893},
  doi = {10.1038/s41566-020-0658-1},
  urldate = {2025-08-11},
  langid = {english}
}

@article{maObservationTunablePolarization2025,
  title = {Observation of Tunable Polarization States Evolution in Bound States in the Continuum Based on Terahertz Metasurfaces},
  author = {Ma, Yumeng and Deng, Fangze and Ma, Ke and Hou, Xiang and Han, Zhihua and Li, Yuchao and Cheng, Keke and Shao, Yansheng and Wang, Chenglong and Liu, Meng and Zhang, Huiyun and Zhang, Yuping},
  year = {2025},
  month = aug,
  journal = {Journal of Applied Physics},
  volume = {138},
  number = {5},
  pages = {053101},
  issn = {0021-8979, 1089-7550},
  doi = {10.1063/5.0282994},
  urldate = {2025-08-08},
  langid = {english}
}

@article{zhaoMagneticallyInducedTopological2025,
  title = {Magnetically {{Induced Topological Evolutions}} of {{Exceptional Points}} in {{Photonic Bands}}},
  author = {Zhao, Xingqi and Wang, Jiajun and Liu, Wenzhe and Shi, Lei and Zi, Jian},
  year = {2025},
  month = jul,
  journal = {Physical Review Letters},
  volume = {135},
  number = {4},
  pages = {046203},
  issn = {0031-9007, 1079-7114},
  doi = {10.1103/wv2n-51qg},
  urldate = {2025-08-11},
  langid = {english}
}

@article{tunableMO2025,
  title = {Tunable Magneto-Optical Accidental Bound States in the Continuum with Intrinsic Chirality and Nonreciprocal Transmission},
  author = {Zhang, Rui and Li, Xiao-Chun and Liu, Qing Huo},
  year = {2025},
  month = jul,
  journal = {Physical Review B},
  volume = {112},
  number = {1},
  pages = {014417},
  issn = {2469-9950, 2469-9969},
  doi = {10.1103/zgm2-nh53},
  urldate = {2025-08-11},
  langid = {english}
}

@article{TwistedMoirePhotonic,
  title = {Twisted Moir{\'e} Photonic Crystal Enabled Optical Vortex Generation through Bound States in the Continuum},
  author = {Zhang, Tiancheng and Dong, Kaichen and Li, Jiachen and Meng, Fanhao and Li, Jingang and Munagavalasa, Sai and Grigoropoulos, Costas P. and Wu, Junqiao and Yao, Jie},
  year = {2023},
  month = sep,
  journal = {Nature Communications},
  volume = {14},
  number = {1},
  pages = {6014},
  issn = {2041-1723},
  doi = {10.1038/s41467-023-41068-1},
  urldate = {2025-08-13},
  langid = {english}
}

@article{GaoACS,
author = {Gao, Siyuan and Liu, Tianji and Iwamoto, Satoshi and Ota, Yasutomo},
title = {Design of Ultrathin Faraday Rotators Based on All-Dielectric Magneto-Optical Metasurfaces at the Telecommunication Band},
journal = {ACS Photonics},
volume = {12},
number = {12},
pages = {6745-6751},
year = {2025},
doi = {10.1021/acsphotonics.5c01826},
URL = {   
        https://doi.org/10.1021/acsphotonics.5c01826
},
eprint = { 
        https://doi.org/10.1021/acsphotonics.5c01826
}
}

@article{PRL-MSPolarization,
  title = {Metasurface Polarization Optics: Independent Phase Control of Arbitrary Orthogonal States of Polarization},
  author = {Balthasar Mueller, J. P. and Rubin, Noah A. and Devlin, Robert C. and Groever, Benedikt and Capasso, Federico},
  journal = {Phys. Rev. Lett.},
  volume = {118},
  issue = {11},
  pages = {113901},
  numpages = {5},
  year = {2017},
  month = {Mar},
  publisher = {American Physical Society},
  doi = {10.1103/PhysRevLett.118.113901},
  url = {https://link.aps.org/doi/10.1103/PhysRevLett.118.113901}
}

@article{MOBICBack-photoniX-anisotropic-review,
	title = {Optical polarization manipulations with anisotropic nanostructures},
	volume = {5},
	issn = {2662-1991},
	url = {https://photoniX.springeropen.com/articles/10.1186/s43074-024-00143-6},
	doi = {10.1186/s43074-024-00143-6},
	language = {en},
	number = {1},
	urldate = {2026-01-29},
	journal = {PhotoniX},
	author = {Li, Zhancheng and Liu, Wenwei and Zhang, Yuebian and Cheng, Hua and Zhang, Shuang and Chen, Shuqi},
	month = oct,
	year = {2024},
	pages = {30},
}

@article{MOBICBack-LSc-MS-Pol,
	title = {Arbitrary polarization conversion dichroism metasurfaces for all-in-one full {Poincaré} sphere polarizers},
	volume = {10},
	issn = {2047-7538},
	url = {https://www.nature.com/articles/s41377-021-00468-y},
	doi = {10.1038/s41377-021-00468-y},
	language = {en},
	number = {1},
	urldate = {2026-01-29},
	journal = {Light: Science \& Applications},
	author = {Wang, Shuai and Deng, Zi-Lan and Wang, Yujie and Zhou, Qingbin and Wang, Xiaolei and Cao, Yaoyu and Guan, Bai-Ou and Xiao, Shumin and Li, Xiangping},
	month = jan,
	year = {2021},
	pages = {24},
}

@article{MOBICBack-LPR-MS-Pol,
	title = {Arbitrary {Achromatic} {Polarization} {Control} with {Reconfigurable} {Metasurface} {Systems}},
	volume = {17},
	issn = {1863-8880, 1863-8899},
	url = {https://onlinelibrary.wiley.com/doi/10.1002/lpor.202200926},
	doi = {10.1002/lpor.202200926},
	language = {en},
	number = {7},
	urldate = {2026-01-29},
	journal = {Laser \& Photonics Reviews},
	author = {Wang, Evan W. and Yu, Shang‐Jie and Phan, Thaibao and Dhuey, Scott and Fan, Jonathan A.},
	month = jul,
	year = {2023},
	pages = {2200926},
}

@article{MOBICBack-LSc-review,
	title = {Optical manipulation from the microscale to the nanoscale: fundamentals, advances and prospects},
	volume = {6},
	issn = {2047-7538},
	shorttitle = {Optical manipulation from the microscale to the nanoscale},
	url = {https://www.nature.com/articles/lsa201739},
	doi = {10.1038/lsa.2017.39},
	language = {en},
	number = {9},
	urldate = {2026-01-29},
	journal = {Light: Science \& Applications},
	author = {Gao, Dongliang and Ding, Weiqiang and Nieto-Vesperinas, Manuel and Ding, Xumin and Rahman, Mahdy and Zhang, Tianhang and Lim, ChweeTeck and Qiu, Cheng-Wei},
	month = mar,
	year = {2017},
	pages = {e17039--e17039},
}

@article{MOBICBack-PRL-signal,
	title = {Nonlocal {Metasurfaces} for {Optical} {Signal} {Processing}},
	volume = {121},
	issn = {0031-9007, 1079-7114},
	url = {https://link.aps.org/doi/10.1103/PhysRevLett.121.173004},
	doi = {10.1103/PhysRevLett.121.173004},
	language = {en},
	number = {17},
	urldate = {2026-01-29},
	journal = {Physical Review Letters},
	author = {Kwon, Hoyeong and Sounas, Dimitrios and Cordaro, Andrea and Polman, Albert and Alù, Andrea},
	month = oct,
	year = {2018},
	pages = {173004},
}

@article{MOBICBack-IEEE-review,
	title = {Advances on {Exploiting} {Polarization} in {Wireless} {Communications}: {Channels}, {Technologies}, and {Applications}},
	volume = {19},
	copyright = {https://ieeexplore.ieee.org/Xplorehelp/downloads/license-information/IEEE.html},
	issn = {1553-877X},
	shorttitle = {Advances on {Exploiting} {Polarization} in {Wireless} {Communications}},
	url = {http://ieeexplore.ieee.org/document/7562438/},
	doi = {10.1109/COMST.2016.2606639},
	language = {en},
	number = {1},
	urldate = {2026-01-29},
	journal = {IEEE Communications Surveys \& Tutorials},
	author = {Guo, Caili and Liu, Fangfang and Chen, Shuo and Feng, Chunyan and Zeng, Zhimin},
	year = {2017},
	pages = {125--166},
}

@article{MOBICBack-Science-sensing,
	title = {Optical polarization–based seismic and water wave sensing on transoceanic cables},
	volume = {371},
	issn = {0036-8075, 1095-9203},
	url = {https://www.science.org/doi/10.1126/science.abe6648},
	doi = {10.1126/science.abe6648},
	language = {en},
	number = {6532},
	urldate = {2026-01-29},
	journal = {Science},
	author = {Zhan, Zhongwen and Cantono, Mattia and Kamalov, Valey and Mecozzi, Antonio and Müller, Rafael and Yin, Shuang and Castellanos, Jorge C.},
	month = feb,
	year = {2021},
	pages = {931--936},
}

@article{MOBICBack-PRL-ecoding,
	title = {Atomic-{Scale} {On}-{Demand} {Photon} {Polarization} {Manipulation} with {High} {Efficiency} for {Integrated} {Photonic} {Chips}},
	volume = {134},
	issn = {0031-9007, 1079-7114},
	url = {https://link.aps.org/doi/10.1103/PhysRevLett.134.083601},
	doi = {10.1103/PhysRevLett.134.083601},
	language = {en},
	number = {8},
	urldate = {2026-01-29},
	journal = {Physical Review Letters},
	author = {Lu, Yunning and Liao, Zeyang and Wang, Xue-Hua},
	month = feb,
	year = {2025},
	pages = {083601},
}

@article{MOBICBack-LPR-communication,
	title = {On‐{Chip} {Polarization}‐ and {Frequency}‐{Division} {Demultiplexing} for {Multidimensional} {Terahertz} {Communication}},
	volume = {16},
	issn = {1863-8880, 1863-8899},
	url = {https://onlinelibrary.wiley.com/doi/10.1002/lpor.202200136},
	doi = {10.1002/lpor.202200136},
	language = {en},
	number = {10},
	urldate = {2026-01-29},
	journal = {Laser \& Photonics Reviews},
	author = {Deng, Wentao and Chen, Liao and Zhang, Hongqi and Wang, Shiwei and Lu, Zijie and Liu, Siqi and Yang, Zuoming and Wang, Ziwei and Yuan, Shixing and Wang, Yilun and Wang, Ruolan and Yu, Yu and Wu, Xiaojun and Yu, Xianbin and Zhang, Xinliang},
	month = oct,
	year = {2022},
	pages = {2200136},
}

@article{MOBICBack-PRL-quantum-key,
	title = {Robust {Polarization}-{Based} {Quantum} {Key} {Distribution} over a {Collective}-{Noise} {Channel}},
	volume = {92},
	copyright = {http://link.aps.org/licenses/aps-default-license},
	issn = {0031-9007, 1079-7114},
	url = {https://link.aps.org/doi/10.1103/PhysRevLett.92.017901},
	doi = {10.1103/PhysRevLett.92.017901},
	language = {en},
	number = {1},
	urldate = {2026-01-29},
	journal = {Physical Review Letters},
	author = {Boileau, J.-C. and Gottesman, D. and Laflamme, R. and Poulin, D. and Spekkens, R. W.},
	month = jan,
	year = {2004},
	pages = {017901},
}

@article{MOBICBack-NC-pol-encode,
	title = {Three-dimensional orientation-unlimited polarization encryption by a single optically configured vectorial beam},
	volume = {3},
	issn = {2041-1723},
	url = {https://www.nature.com/articles/ncomms2006},
	doi = {10.1038/ncomms2006},
	language = {en},
	number = {1},
	urldate = {2026-01-29},
	journal = {Nature Communications},
	author = {Li, Xiangping and Lan, Tzu-Hsiang and Tien, Chung-Hao and Gu, Min},
	month = aug,
	year = {2012},
	pages = {998},
}

@article{gxgygz,
	title = {Magneto-optics of multilayers with arbitrary magnetization directions},
	volume = {43},
	copyright = {http://link.aps.org/licenses/aps-default-license},
	issn = {0163-1829, 1095-3795},
	url = {https://link.aps.org/doi/10.1103/PhysRevB.43.6423},
	doi = {10.1103/PhysRevB.43.6423},
	language = {en},
	number = {8},
	urldate = {2026-02-09},
	journal = {Physical Review B},
	author = {Zak, J. and Moog, E. R. and Liu, C. and Bader, S. D.},
	month = mar,
	year = {1991},
	pages = {6423--6429},
}

@article{tunable-graphene,
  title = {Dynamically tunable and ultrastable plasmonic bound states in the continuum in bilayer graphene metagratings},
  author = {Huang, Jiali and Qing, Guizi and Zhang, Di and Zhai, Xiang and Peng, Jun and Xia, Sheng-Xuan},
  journal = {Phys. Rev. B},
  volume = {112},
  issue = {20},
  pages = {205421},
  numpages = {10},
  year = {2025},
  month = {Nov},
  publisher = {American Physical Society},
  doi = {10.1103/2yc3-xm7g},
  url = {https://link.aps.org/doi/10.1103/2yc3-xm7g}
}

@article{tunableBICactive,
author = {Haoqi Luo and Liangliang Liu and Junyu Zhang and Qing Ye and Yihua Hu and Fengya Lu},
journal = {Opt. Express},
number = {2},
pages = {1703--1723},
publisher = {Optica Publishing Group},
title = {Tunable bound states in the continuum with loss compatibility},
volume = {33},
month = {Jan},
year = {2025},
url = {https://opg.optica.org/oe/abstract.cfm?URI=oe-33-2-1703},
doi = {10.1364/OE.547894},
}

% Full bibliography added automatically for Optics Letters submissions; the following line will simply be ignored if submitting to other journals.
% Note that this extra page will not count against page length
\bibliographyfullrefs{sample}

%Manual citation list
%\begin{thebibliography}{1}
%\bibitem{Zhang:14}
%Y.~Zhang, S.~Qiao, L.~Sun, Q.~W. Shi, W.~Huang, %L.~Li, and Z.~Yang,
 % \enquote{Photoinduced active terahertz metamaterials with nanostructured
  %vanadium dioxide film deposited by sol-gel method,} Opt. Express \textbf{22},
  %11070--11078 (2014).
%\end{thebibliography}

% Please include bios and photos of all authors for aop articles
% \ifthenelse{\equal{\journalref}{aop}}{%
% \section*{Author Biographies}
% \begingroup
% \setlength\intextsep{0pt}
% \begin{minipage}[t][6.3cm][t]{1.0\textwidth} % Adjust height [6.3cm] as required for separation of bio photos.
%   \begin{wrapfigure}{L}{0.25\textwidth}
%     \includegraphics[width=0.25\textwidth]{john_smith.eps}
%   \end{wrapfigure}
%   \noindent
%   {\bfseries John Smith} received his BSc (Mathematics) in 2000 from The University of Maryland. His research interests include lasers and optics.
% \end{minipage}
% \begin{minipage}{1.0\textwidth}
%   \begin{wrapfigure}{L}{0.25\textwidth}
%     \includegraphics[width=0.25\textwidth]{alice_smith.eps}
%   \end{wrapfigure}
%   \noindent
%   {\bfseries Alice Smith} also received her BSc (Mathematics) in 2000 from The University of Maryland. Her research interests also include lasers and optics.
% \end{minipage}
% \endgroup
% }{}

\end{document}